\documentclass[aps,floatfix]{revtex4}

\begin{document}

\title{Higher-dimensional Embedding of Four-dimensional Hypersurfaces: Chronological Development of Co-dimension-1 Embedding}
\author{Sarbari Guha \cite{email}}
\affiliation{Department of Physics, St. Xavier's College, Calcutta, India}
\date{\today}

\begin{abstract}
Here we present an overview of the work done over the years on the embedding of hypersurfaces in a space of higher dimension, with particular reference to the embedding of four-dimensional hypersurfaces in five-dimensional space-times. The concept of embedding was developed by geometers starting from Gauss. Applications in physics started with the embedding of the Schwarzschild metric and through the works of Kaluza and Klein. The program on the embedding of four-dimensional hypersurfaces in five-dimensional spacetimes gathered strength with the formulation of the induced matter theory by Wesson and his collaborators. Close on its heels followed the works on the braneworld scenario. Simultaneously, the mathematical aspects of the embedding of four-dimensional hypersurfaces in five-dimensional spacetimes were developed as an application of the Campbell-Magaard theorem. In this paper we have tried to have a panoramic view of the major works done in these fields and have tried to focus on the recent trends in this context.
\end{abstract}

\maketitle\textbf{INTRODUCTION}

\bigskip

Einstein had dreamt of a purely geometric description of all physical interactions including the description of matter fields. This is the basic idea of the unified field theory. Various extensions of Einstein's approach to extract matter purely from geometry have been made. Such theories are mostly developed in more than four-dimensions embracing the geometrical concept of embedding.

\bigskip

\maketitle\textbf{HISTORICAL BACKGROUND}

\bigskip

The problem of isometric embedding had been considered by the earlier geometers starting with Gauss \cite{gauss}. Riemann had introduced the concept of an abstract manifold whose geometrical properties were characterized solely by its intrinsic properties, without any reference to a higher dimensional embedding space \cite{riemann}.  Eisenhart \cite{eisenhart} devoted a large portion of his famous book to the description of Riemann spaces with added dimensionality. However, from the point of view of physics, the embedding of general relativistic spacetimes started with the effort to embed the Schwarzschild metric. Today we know that any intrinsically defined Riemannian manifold can be isometrically embedded both globally as well as locally in a Euclidean (or Minkowski) space of appropriate dimension and signature. However global embeddings of Riemannian spaces (or space-times) are much more complicated and far less studied than local embeddings \cite{rtzgrg}.

\bigskip

Embedding theories are naturally subject to theorems on differential geometry, leading to a deeper understanding of the geometry of the lower-dimensional world and the corresponding higher dimensional embedding space. In this regard, an important theorem due to Campbell and Magaard, along with its extended versions, promises to throw new light on the mathematical structure of the non-compactified theories of gravity \cite{wesson}-\cite{rs2}. The well-known Janet-Cartan theorem \cite{janet} on local isometric embeddings of Riemannian manifolds states that: \textit{if the embedding space is flat, then the minimum number of extra dimensions needed to analytically embed an n-dimensional Riemannian manifold is d},  with $0 \leq d \leq n(n - 1)/2$. The innovation of the Campbell-Magaard theorem \cite{cm}, is that the number of extra dimensions falls drastically to $d=1$, when the embedding manifold is allowed to be Ricci-flat (vacuum solution of Einstein's equations), instead of Riemann-flat \cite{rtzgrg}. This theorem and its extensionss hold irrespective of the signature, and is therefore applicable to classical relativity as well as Euclidean quantum gravity. Campbell's theorem also brings forth some interesting questions, one of which concerns the physical meaning of curvature in 5-dimensional space. Embedding geometric manifolds in higher dimensional spaces can be a useful tool for studying global properties of these manifolds.

\bigskip

The results on isometric embeddings opened new perspectives in the realms of theoretical physics, particularly in the studies of the gravitational field. On one hand it serves as a mathematical tool to construct and classify solutions of General Relativity. On the other hand, it allows to introduce new variables which are better suited for an appropriate description of quantum mechanical phenomena \cite{seminar}. The theory of elementary particle physics developed in terms of a flat space (commonly called the Minkowski space) and its symmetry properties, which is the symmetry of the Lorentz group. Attempts to describe the properties of particles in curved space-time led to a renewed interest in the embedding scheme. Attempts have been made to identify the internal symmetries of the particles with the symmetries of the generalized space-time curvature. The mathematical aspect of the problem is then to obtain information on the embedding class of various relativistic metrics, which is the minimal number of extra dimensions necessary to satisfy the Gauss-Codazzi-Ricci equations. Many workers have contributed to the studies of the mathematical and physical aspects of embedding \cite{jmp}. By the very nature of the embedding process the extra dimensions find their origin in the presence of gravity. Moreover, if the embedding is required to be a minimal one, then the number of extra dimensions of the higher-dimensional space is fixed.

\bigskip

\maketitle\textbf{RECENT RESEARCHES}

\bigskip

Recently there has been a renewed interest in theories of gravity with extra dimensions. Historically, investigations along this direction was started by Kaluza and Klein \cite{kk} to interpret the fundamental interactions, where the fifth dimension was utilized as the source of new degrees of freedom to be associated with the electromagnetic field. Extra dimensions also arose naturally in superstring and supergravity theories from the necessity to ensure a non-anomalous quantum behavior and the attempts to construct a fundamental theory of physical interactions. These extra dimensions might have played an important role in the evolution of the early universe.  An immediate task within this higher-dimensional framework is to explore how to explain the 4-dimensionality of the observed universe. However, these extra dimensions were constrained to be compactified to very small scales(of the order of Planck length), since they remained undetected in experiments at energy levels within the TeV scale. Even then it would be necessary to find another explanation for the non-observability of the extra dimensions, as for example, by breaking the translational symmetry along the extra dimension \cite{rbvspv}.

\bigskip

The success of the Kaluza-Klein theory in particle physics led several workers to construct higher-dimensional cosmological models \cite{visser}-\cite{rippl}. The problem with the cosmological models in more than four dimensions was that the form of the energy-momentum tensor was unsure, making the models physically ambiguous. This problem was circumvented in Wesson's IMT model \cite{wesson}, where the energy density and pressure of the $(3+1)$-dimensional energy-momentum tensor arose directly from the extra components of the $(4+1)$-dimensional Einstein tensor for vacuum. The basic ideas of this theory was developed by a number of researchers\cite{imt1}-\cite{imt12}. The difference between the Kaluza-Klein scheme and Wesson's scheme was that the latter did not attempt to geometrize the electromagnetic field and there was no assumption concerning the compactness of the extra dimension. The motive behind the assumption of the existence of a large extra dimension was to achieve the unification of matter and geometry, in the sense that the properties of matter could be obtained as a consequence of the extra dimensions. Thus, in principle, the physics of the $(3+1)$-dimensional cosmologies could be recovered from the geometry of the $(4+1)$-dimensional vacuum field equations. The new solutions in five-dimensions were found to have significant applications, although for the explanation of the gauge properties of elementary particles, something more complicated than five-dimensional Riemannian geometry is felt necessary. Rippl et. al. \cite{rippl} used this approach to go from the four-dimensional case to lower dimensional gravity, the latter being easily quantizable than Einstein's gravity. It was found that a class of solutions of the five-dimensional equations of GR that are static, spherically-symmetric metric in ordinary three-dimensional space and which depend on 3 parameters, represent shells of matter around a local centre in the induced matter interpretation. It was observed that such solutions could have applications in astrophysics and particle physics \cite{wesliuplb}. As the search for a grand unification continues, the overlapping between the researches in the mathematical aspects of embedding and those in the field of superstring and supergravity is becoming more and more prominent.

\bigskip

The works of Arkani-Hamed and others \cite{aadd}\cite{add}have provided a new framework for solving the electroweak hierarchy problem i.e. the presence of the large gap between the scale of gravity and the scale of other interactions.. The gravitational and gauge interactions unite at the electroweak scale and the observed weakness of gravity at long distances is due to the existence of large new spatial dimensions. General relativity fails to describe gravity at high enough energies and has to be replaced by a quantum theory of gravity (that removes the classical singularities predicted by GR), picking up significant corrections as the fundamental energy scale is approached \cite{lrr}. This idea led to the so-called 'braneworld scenario' where the ordinary standard model matter and non-gravitational fields are confined by some trapping mechanism to the four-dimensional universe constituting the D3-branes (four-dimensional timelike hypersurfaces i.e. submanifold) which are embedded in a $(4 + n)$-dimensional manifold referred to as the 'bulk' (n being the number of extra dimensions). At low energies, gravity is localized at the brane along with particles and general relativity is recovered, but at high energies gravity "leaks" into the higher-dimensional bulk (i.e. the entire space-time) and is propagated therein via the equivalence principle, so that only a part of it is felt in our 4-dimensions. The strong constraints on the size of extra dimensions as in the Kaluza-Klein models, is therefore relaxed \cite{add}, leading to a low fundamental Planck scale (as low as the TeV scale), thereby involving many phenomenological consequences. According to Randall and Sundrum \cite{rs1}\cite{rs2}, there is only a single extra dimension, which is curved  (warped) and need not be compact.  This, in principle, seeks the equivalence between the IMT and brane-world theories \cite{poncempla}. The main role of extra dimension in such theories is to provide extra degrees of freedom (both scalar and vector degrees of freedom) on the brane. However, unlike the RS models, where the matter field is confined to the fixed brane, in the IMT model, there is no mechanism to confine induced matter field exactly on a specific brane. Furthermore, the rest mass of particles vary from point to point in the spacetime, in agreement with Mach's idea \cite{jalalyaz}.

\bigskip

The novel feature of the RS models compared to previous higher-dimensional models is that, here the observable three spatial dimensions are protected from the large extra dimension (at low energies) by curvature rather than straightforward compactification. In the RS braneworlds, gravity is localized at the brane due to curvature of the bulk (warped compactification) and not due to compactification of the fifth dimension . At low energies, the negative bulk cosmological constant prevents gravity from leaking into the extra dimension, giving rise to a warping effect, thereby squeezing the gravitational field closer to the brane. The exponential warp factor reflects the confining role of the bulk cosmological constant. The curvature radius determines the magnitude of the Riemann tensor. In the bulk, the metric is a solution of the Einstein equations for a non-compact ambient space, which in this case, is a double-sided ($Z_{2}$-symmetric) five-dimensional anti-de Sitter spacetime ($AdS_{5}$), whose boundary is the brane. The $Z_{2}$-symmetry about the brane at $y=0$ is incorporated through the $|y|$ term. The brane is a flat Minkowski spacetime, with self-gravity in the form of brane tension, thereby compensating for the bulk cosmological constant, so that the quadratic leading order due to the brane contribution would be cancelled, and one could recover the linearized Einstein equations in the general relativistic limit \cite{CHR}. A thorough analysis  of this feature was done by a number of workers, who also calculated exact corrections to the Newton's potential \cite{garriga}. The brane tension arises from the non-linearity of the dynamical equations in the energy-momentum tensor. Randall and Sundrum have proposed two models, of which the first one has a compactified fifth dimension, bounded by two branes with positive and negative tension, the visible universe being modelled by the negative tension brane. Although in the second model, a single flat brane is embedded into an AdS bulk in a mirror-symmetric way with a non-compact fifth dimension, later generalizations have evolved into considering curved branes, embedded both symmetrically and asymmetrically into a bulk characterized by both Weyl and Ricci curvatures. Motivated by string/ M-theory, models where the bulk can support fields outside the standard model (e. g. moduli or dilatonic scalar fields or a radiation of quantum origin), have been proposed.

\bigskip

In the RSII model \cite{rs2}, the brane is infinitely thin along the extra dimension. The brane tension renders the first derivative of the metric to be discontinuous at the brane. The thin brane approximation is valid as long as the energy scales of the model are much smaller than the energy scale related to the inverse thickness of the brane so that we may not neglect the thickness of the brane at the fundamental scale. A thick brane is constructed from a real scalar field and are regular domain wall solutions having a finite thickness, thereby giving rise to well-behaved differentiable metrics \cite{bonjour}. The equation of motion of the brane is obtained by writing down the proper junction conditions on the boundaries of the brane with the bulk, and the effect of the thickness on its motion is determined \cite{KM}. The motion of spherical and planar thick domain walls was first treated by Garfinkle and Gregory \cite{garfinkle}. The confinement of gravity is also valid in the thick brane version of RSII models \cite{chamblin}. A strong motivation for considering such brane models comes from the strongly coupled string theories. For all such theories, an essential issue concerns the cosmological evolution of our universe. Cosmological models with a Robertson-Walker brane are found to tend at late times to the standard big bang model, representing the observed universe \cite{kaloper}\cite{bdl}. It has been demonstrated \cite{bdl}\cite{deruelle} that the equations governing the cosmological evolution of the brane will be generically different from those corresponding to the analogous Friedmann equations of standard cosmology. Essentially the difference lies in the fact that the energy density of the brane appears in \emph{quadratic form} in the brane equations although it is \emph{linear} in the standard cosmology. This leads to unconventional cosmological evolution in some cases \cite{cline}. Dynamical evolution is possible when the brane tension does not exactly compensate the bulk cosmological constant \cite{kaloper1}. Perturbations of such cosmological models have been proposed and analysed to calculate the CMB anisotropies predicted by brane worlds \cite{mukhoyama}\cite{langlois}\cite{deruelle1}. Extensive work has been carried out in the theory of gauge-invariant perturbations in brane-world cosmology \cite{lrr1}. In the DGP models \cite{dgp}, modification of braneworld gravity is proposed to explain the accelerating expansion of the universe. The cosmological solution of the DGP theory exhibits self-acceleration on the brane at late times without the need to include any exotic cosmological fluid, describing a universe that is accelerating beyond the cross-over scale \cite{deffayet}.

\bigskip

For the embedding of a single Minkowski 3-brane with non-zero spatial curvature into a 5-dimensional non-compactified spacetime bulk \cite{mannheim}, a five-dimensional bulk cosmological constant is not able to produce the exponential suppression of the geometry thought necessary to localize gravity to the brane. By use of the Einstein equations, it has been shown that \cite{constcurv} the embedding of a surface in a flat space of co-dimension one imposes the restriction that the surface has a constant curvature, if its dimension $n > 2$. Therefore to embed braneworlds which do not have constant curvature, we need at least two extra dimensions \cite{andrianopoli}.

\bigskip

Exact solutions of the global Einstein's equations with a bulk cosmological constant exist only when the space-time is static \cite{rs1}. The effective equations for gravity in four dimensions (the brane field equations, BEFE) were obtained by Shiromizu et al \cite{shiromizu}using Israel's junction conditions \cite{israel} and $Z_{2}$-symmetry for the bulk space time, in which the brane is embedded. Cosmological solutions have been studied \cite{ida}.  However, these BEFE do not constitute a closed system, as they contain an 'electric' Weyl tensor term, which can only be specified in terms of the five-dimensional bulk. Further, these equations contain a term quadratic in the energy-momentum tensor of the brane, which can have important consequences for the evolution of models depicting the universe in its early epochs, where we approach the high energy limit \cite{bdl} \cite{maartens}. This is the high energy inflation constraint, which originates from the junction conditions following the assumption of reflection symmetry about the thin brane embedded in the bulk, and is not consistent with the regularity conditions of the braneworld. The Friedmann equation on the brane in these models is modified by both the high-energy matter terms and a term carrying the non-local bulk effects onto the brane. The matter fields confined to the brane make up the brane energy-momentum tensor. Although it is usually assumed that the space-time is exactly anti-de Sitter in the absence of a brane, this is not necessarily the case. The brane-free bulk metric can be any solution of the Einstein's equations on the bulk, including non-conformally flat solutions. The projection of the bulk Weyl tensor onto the brane gives rise to an additional term in the Friedmann equations, which is the most important contribution of the bulk geometry to the brane. This supplies a geometric perfect fluid, called the Weyl fluid, whose nature is governed by the contents of the bulk, empty or non-empty. This entity can play an important role in the cosmological dynamics and perturbations. Specifically, it can play a crucial role in the late time evolution of the universe as a geometric candidate for dark matter \cite{pal}.

\bigskip

Some observational constraints on the braneworlds based on predictions from the specific models in five-dimensions, have recently been reported, both on local and cosmological scales. Particularly, the interference of linear gravitational waves generated by the bulk geometry over the quadrupole formula for the binary pulsar PSR$1913+16$, predicts an error of about 20 percent, against an observational error of just $0.1$ percent \cite{durrer}. Here the authors show that in the gravitational models with compact extra dimensions, the linearly perturbed Einstein equations do not agree with the observations. There are different models that lead to a modified Friedmann's equation depending on the square of the energy density, thereby slowing down the high energy inflation, in disagreement with the recent data from the WMAP/ SDSS/ 2DF \cite{tsujikawa}. A general solution of this problem based on differential geometry suggests a stable bulk geometry, whose existence requires higher dimensions \cite{maia1}. The presence of the high energy inflation constraint tells us that $Z_{2}$-symmetry and orbifold compactification must not be present on the braneworld and possibly the braneworld program requires more than five dimensions, although there are other merits of choosing $Z_{2}$-symmetry, as for example, the coincidence of the geodesics of test particles in the brane with the geodesics in the bulk.

\bigskip

The dark energy component of the universe not only explains the observed accelerating expansion of the universe but also provides the remaining piece of information connecting the inflationary flatness prediction with the current astronomical observations. A geometrical investigation of the dark energy component within the context of braneworld theory without mirror symmetry leads to its geometrical interpretation as a warp in the universe given by the extrinsic curvature of the brane. The cosmology involves the standard Friedmann universe embedded in a 5-dimensional and constant curvature bulk without any a priori junction condition between the brane and the bulk \cite{maia2}.

\bigskip

Running parallel to these researches there has been quite a substantial investigation in the geometrical aspects of embedding theories of space-time \cite{dahia1}. The Campbell-Magaard theorem has been extended to more general contexts where the higher dimensional space is sourced by a cosmological constant or by a scalar field and even by an arbitrary non-degenerate Ricci tensor \cite{anderson1}\cite{dahia1}. It happens that for the case where the embedding space is a vacuum bulk having a source which is singular i.e. corresponds to a Dirac delta function supported by a distribution of matter concentrated in an n-dimensional hypersurface (thin shell), the extrinsic curvature of the hypersurface is not continuous . The discontinuity of the extrinsic curvature is proportional to the energy-momentum tensor of the hypersurface thereby requiring two different embeddings, one for each side of the hypersurface, matched at the hypersurface by junction conditions. Here, the energy content of the space-time (i.e. the brane) represented by its energy-momentum tensor does not necessarily determine the curvature of the space-time, the only required relation between the energy-momentum tensor and the metric tensor being the conservation equation. As a result it is possible to embed Minkowski space-time that has an extrinsic curvature representing dust with a uniform density, in a 5-dimensional bulk.

\bigskip

The intense activities in these higher dimensional scenarios have strongly motivated the search for a proper geometrical formulation necessary to support the physical theories so as to match them with astronomical observations. For example, the search for geometrical mechanisms which could allow confinement of particles has led to the investigation of the warped product spaces and their geometrical properties \cite{felder}. A recent work on the dynamics of particles in the context of the IMT theory, has shown that the anomalous acceleration of particles can be dealt with either by the foliation approach or by the embedding approach \cite{romero}. Originally most braneworld and higher dimensional studies concentrated on codimension one i.e. $(4 + 1)$-dimensional models. However, study of higher codimensional models is called for in view of the predictions of the string theory \cite{chatillon}. Given that the observed macroscopic world is 4-dimensional, the question that remains is that how closely the corresponding apparent 4-dimensional worlds resemble the observed world and how do the observations made in their apparent worlds can test and set constraints on the full theories \cite{anderson2}. It is therefore necessary to verify the extent to which these theories satisfy the geodesic postulates concerning the motion of test particles and light rays i.e. the behaviour of geodesics in these models must be determined. In particular, if possible, the relation between the geodesics of the higher-dimensional space and those belonging to the hypersurface has to be established.

\bigskip

The geodesics of a codimension one submanifold are not generically among those of the manifold. For a general thin asymmetric braneworld, bulk and brane geodesics does not coincide in general, so that bulk-propagating and brane-confined matter test bodies undergo distinct free fall, which, in principle violates the equivalence principle and there is loss or gain of the bulk species matter on the brane. The $Z_{2}$-symmetric braneworlds have their apparent and bulk geodesics coinciding on the brane and are considerably more general than totally-geodesic submanifolds. The study of classical geodesic motions of non-zero rest mass test particles and photons in 5-dimensional warped product spaces indicate that it is possible to obtain a general picture of these motions, using the natural decoupling occurring between motions in the brane and motion in the fifth dimension in case of such spaces \cite{dahia3}. Consequently, phase space analysis of the possible confinement of particles and photons to hypersurfaces in 5-dimensional warped product spaces, could be made, revealing a novel type of quasi confinement which is oscillatory and neutrally stable. The corresponding warped product geometry is locally $Z_{2}$-symmetric. Such a confinement does not require the presence of usual brane-type confinement mechanisms and is purely due to the classical gravitational effects. For the $(3 + 1 + n)$-dimensional warped product spaces \cite{dahia4}, these motions are not constrained by the value of the extrinsic curvature. The general conditions for the confinement of geodesics in the case of pseudo-Riemannian manifolds have been obtained and the stability conditions for such confinements are established.

\bigskip

In case of the RSII-type thick brane, a test particle following a geodesic motion does not stay bound to the 4-dimensional spacetime if its motion is transversally perturbed, indicating that gravity alone cannot be responsible for the confinement of test particles to this kind of brane. Assuming a direct interaction between test particles and a scalar field (just as the Yukawa-like interaction between fermions and scalar field), it has been found \cite{dahia5} that a stable confinement in a domain wall is possible. Due to this interaction, light test particles remain bound to the brane for small transversal perturbations. In case of heavier particles, such perturbations impart so large an inertia that the interaction with the scalar field is not strong enough to prevent their motion into the extra dimension. In case of the Schwarzschild thick brane, in the low energy regime, the motion of test particles does not reveal the existence of an extra dimension neither by direct observation of the oscillatory motion in the fifth dimension, nor by deviation of the motion in the 4-dimensional ordinary space-time.

\bigskip

Because of the complexity of the full brane-bulk system of equations, very often the non-Einsteinian terms in the BEFE (referred to as the 'bulk' term or Weyl term) is discarded during the studies of the braneworld models retaining only the 'brane' terms. However, such truncations often lead to different imprints on the residual 'braneworld physics' and so all such truncations are unsatisfactory \cite{maartens1}. Hence, a reformulation is done for the full brane-bulk system using geometrical identities \cite{anderson3}.

\bigskip

In general, the geometry of the brane is assumed to be Riemannian to avoid possible conflicts with the well-established theory of general relativity. The geometry of the bulk, however, has seldom been considered, which is generally supposed to be also Riemannian. In an attempt to broaden this scenario, Weyl geometry, which is one of the simplest generalizations of Riemannian geometry, has been taken into consideration as a viable possibility to describe the bulk \cite{israelit}. The correct variant is the Weyl integrable geometry, which is necessary for the successful correlation of relativistic gravitation and elementary particle physics. Therefore Riemannian theorems have been generalized to Weyl geometry and Riemannian space-time is locally and isometrically embedded in a Weyl bulk having the geometry of warped product space \cite{dahia6}. In such models, a Weyl scalar field  that depends only on the extra coordinate has been considered to construct an analogue of quantum confinement.

\bigskip

\maketitle\textbf{CONCLUDING REMARKS}

\bigskip

The concept of embedding was developed by geometers starting from Gauss. Applications in physics started with the embedding of the Schwarzschild metric and was followed by the works of Kaluza and Klein to build up a unified theory of fundamental interactions. The success of the Kaluza-Klein theory in particle physics led several workers to construct higher-dimensional cosmological models. The program on the embedding of four-dimensional hypersurfaces in five-dimensional spacetime gathered strength in the last decade of the previous millenium with the formulation of the induced matter theory by Wesson and his collaborators, to geometrize the electromagnetic field. However, to achieve the unification of matter and geometry, the assumption concerning the compactness of the extra dimension was removed. Extra dimensions occurred naturally in superstring and supergravity theories to ensure a non-anomalous quantum behavior and the attempts to construct a fundamental theory of physical interactions. The works on the braneworld scenario tried to address the hierarchy problem and other issues unsolved in the string theories. Simultaneously, the mathematical aspects of the embedding of four-dimensional hypersurfaces in five-dimensional spacetime were developed as an application of the Campbell-Magaard theorem. In this paper we have tried to have a panoramic view of the major works done in these fields and have tried to focus on the recent trends in this context with reference to the studies on the "Weyl" fluid. However, to keep the length suitable for an easy reading, selected references have been cited. The intention is to lead the reader to some of the notable papers, which can serve as reference to other similar works.
\bigskip

\maketitle\textbf{ACKNOWLEDGEMENTS}

\bigskip

The author thanks the University Grants Commission of the Government of India for financial support. Most of the reference materials were available through her affiliation to IUCAA, INDIA as a visiting associate.


\begin{thebibliography}{}\label{sec:TeXbooks}
\bibitem[*]{email}{Electronic Address: sarbariguha@rediffmail.com,srayguha@yahoo.com}
\bibitem{gauss}K. F. Gauss, \textit{Disquisitiones generales circa superficies curvas}(1827). English translation: \textit{General Investigations of Curved Surfaces of 1827 and 1825} by C. J. Morehead and A. M. Hiltebeitel (New York: Raven Press, 1965) in Karl. Friedrich Gauss Werke, IV (1827).
\bibitem{riemann}B. Riemann, \textit{\"{U}ber die Hypothesen welche der Geometrie zu Grunde liegen} (1854). This thesis was presented on June 10th, 1854, in G\"{o}ttingen. It was first published in Abh. K\"{o}nigl. gesellsch. \textbf{13}, 1 (1868). It was translated into English by W. K. Clifford and published in Nature \textbf{8}, 14 (1873).
\bibitem{eisenhart}L. P. Eisenhart, Riemannian Geometry (Princeton University Press, 1949).
\bibitem{rtzgrg}C. Romero, R. Tavakol and R. Zalaletdinov, Gen. Rel. Grav. \textbf{28}, 365 (1996).
\bibitem{wesson}P. S. Wesson, Space-Time-Matter (World Scientific, Singapore, 1999).
\bibitem{overduin}J. M. Overduin and P. S. wesson, Phys. Rep. \textbf{283}, 302 (1997).
\bibitem{rs1}L. Randall and R. Sundrum, Phys. Rev. Lett. \textbf{83}, 3370 (1999).
\bibitem{rs2}L. Randall and R. Sundrum, Phys. Rev. Lett. \textbf{83}, 4690 (1999).
\bibitem{janet}See Ref. 3 above, p. 143.
\bibitem{cm}J. E. Campbell, A Course of Differential Geometry (Clarendon Press, Oxford, 1926); L. Magaard, \textit{Zur einbettung Riemannscher raume un Eienstein raume und konformeuclidische raume} (Ph.D Thesis, Kiel, 1963).
\bibitem{seminar}I. Robinson and Y, Ne'eman, \textit{Seminar on the Embedding Problem}, Rev. Mod. Phys. \textbf{37}, 201 (1965) and the subsequent individual papers making up the seminar report, namely those by A. Friedman, J. Rosen, R. Penrose, C. Fronsdal, D. W. Joseph and Y. Ne'eman respectively in order of the sequence.
\bibitem{jmp}Innumerable papers have been published from the mid 1960s onwards. Notable among them are those published in J. Math. Phys. Some of those are by C. D. Collinson, \textbf{7}, 608 (1966) and \textbf{9}, 403 (1968); G. H. Goedecke, \textbf{15}, 789 (1974); J. Krause, \textbf{16}, 1090 (1975); C. F. Chyba, \textbf{23}, 1662 (1982); M. D. Maia, \textbf{25}, 2090 (1984); Y. K. Gupta, S. P. Sharma and R. S. Gupta, \textbf{25}, 3510 (1984); I. Ozsv\'{a}th and L. Sapiro, \textbf{28}, 2066 (1987); I. Ozsv\'{a}th, \textbf{29}, 825 (1988) and \textbf{33}, 221 and 229 (1992).
\bibitem{kk}T. Kaluza, \textit{Zum Unit\"{a}tsproblem der Physik}, Sitz. Preuss. Akad. Wiss. Phys. Math. \textbf{K1} 966 (1921). English translation in \textit{Unified field theories of more than four dimensions}, Proc. Int. School of Cosmology and Gravitation (Erice) edited by V. De Sabbata and E. Schmutzer (World Scientific, Singapore, 1983).
\bibitem{rbvspv}V. A. Rubakov and M. E. Shaposhnikov, Phys. Lett. \textbf{125B}, 136 (1983).
\bibitem {visser}M. Visser, Phys. Lett. \textbf{159B}, 22 (1985).
\bibitem{banerjee}See for example A. Banerjee, B. Bhui and S. Chatterjee, Astrophys J. \textbf{358}, 23 (1990) and references therein.
\bibitem{rippl}S. Rippl, C. Romero and R. Tavakol, Class. Quant. Grav. \textbf{12} 2411 (1995).
\bibitem{imt1}P. S. Wesson, Gen. Rel. Grav. \textbf{16}, 193 (1984).
\bibitem{imt2}J. Ponce de Leon, J. Math. Phys. \textbf{20}, 539 (1988).
\bibitem{imt3}P. S. Wesson, Astrophys. J. \textbf{394} 19 (1992); P. S. Wesson and J. Ponce de Leon, J. Math. Phys. \textbf{33}, 3883 (1992).
\bibitem{imt4}P. S. Wesson, Astrophys. J. \textbf{436}, 547 (1994).
\bibitem{imt5}D. J. McManus, J. Math. Phys. \textbf{35} 4889 (1994); A. A. Coley and D. J. McManus, J. Math Phys. \textbf{36}, 335 (1995); G. Abolghasem, A. A. Coley and D. J. McManus, J. Math. Phys. \textbf{37} 361 (1996).
\bibitem{imt6}H. Liu and P. S. Wesson, Int. J. Mod. Phys. D \textbf{3}, 627 (1994).
\bibitem{imt7}P. S. Wesson et. al., Int. J. Mod. Phys. A \textbf{11}, 3247 (1996).
\bibitem{imt8}A. P. Billyard and A. A. Coley, Mod. Phys. Lett. A \textbf{12}, 2121 (1997).
\bibitem{imt9}H. L. Carrion, M. J. Reboucas and A. F. F. Teixeira, J. Math. Phys., \textbf{40} 4011 (1999).
\bibitem{imt10}W. N. Sajko and P. S. Wesson, Mod. Phys. Lett. A \textbf{16}, 627 (2001).
\bibitem{imt11}P. S. Wesson and H. Liu, Int. J. Mod. Phys. D \textbf{10} 905 (2001).
\bibitem{imt12}S. Chatterjee, D. Panigrahi and A. Banerjee, Class. Quant. Grav. \textbf{11} 371 (1994).
\bibitem{wesliuplb}P. S. Wesson and H. Liu, Phys. Lett. B \textbf{432} 266 (1998).
\bibitem{aadd}N. Arkani-Hamed, S. Dimopoulos and G. Dvali, Phys. Lett B \textbf{429}, 263 (1998); I. Antoniadis, N. Arkani-Hamed, S. Dimopoulos, G. Dvali, Phys. Lett B \textbf{436}, 257 (1998).
\bibitem{add}N. Arkani-Hamed, S. Dimopoulos and G. Dvali, Phys. Rev. D \textbf{59}, 086004 (1999).
\bibitem{lrr}Roy Maartens, \textit{Brane-World Gravity}, Living Rev. Relativity, 7, (2004), http://www.livingreviews.org/lrr-2004-7 and references therein.
\bibitem{poncempla}J. Ponce de Leon, Mod. Phys. Lett A \textbf{16}, 2291 (2001).
\bibitem{jalalyaz}S. Jalalzadeh and A. M. Yazdani, gr-qc /0805.3017.
\bibitem{CHR}A. Chamblin, S. W. Hawking and H. S. Reall, Phys. Rev. D \textbf{61} 065007 (2000).
\bibitem{garriga}Some of these references are J. Garriga and T. Tanaka, Phys. Rev. Lett. \textbf{84}, 2778 (2000); C. Csaki, J. Ehrlich, T. J. Hollowood and Y. Shirman, Nucl. Phys. B \textbf{581} 309 (2000).
\bibitem{bonjour}F. Bonjour, C. Charmoussis and R. Gregory, Phys. Rev. D \textbf{62}, 083504 (2000)
\bibitem{KM}S. Khakshournia and R. Mansouri, Gen. Rel. Grav. \textbf{34}, 1847 (2002).
\bibitem{garfinkle}D. Garinkle and R. Gregory, Phys. rev. D \textbf{41}, 1889 (1990).
\bibitem{chamblin}see for example, A. Chamblin and G. W. Gibbons, Phys. Rev. Lett. \textbf{84}, 1090 (2000).
\bibitem{kaloper}N. Kaloper and A. Linde, Phys. Rev. D \textbf{59}, 101303 (1999).
\bibitem{bdl}P. Binetruy, C. Deffayet and D. Langlois, Nucl. Phys. B \textbf{565}, 269 (2000); P. Binetruy, C. Deffayet, U. Ellwanger and D. Langlois, Phys. Lett. B \textbf{477}, 285 (2000).
\bibitem{deruelle}N. Deruelle and T. Dolezel, Phys. Rev. D \textbf{62}, 103502 (2000).
\bibitem{cline}J. M. Cline, C. Grojean and G. Servant, Phys. Rev. Lett. \textbf{83}, 4245 (1999).
\bibitem{kaloper1}N. Kaloper, Phys. Rev. D \textbf{60}, 123506 (1999); hep-th/9905210.
\bibitem{mukhoyama}S. Mukhoyama, Phys. Rev. D \textbf{62}, 064022 (2000);
\bibitem{langlois}D. Langlois, Phys. Rev. D \textbf{62}, 126012 (2000);
\bibitem{deruelle1}N. Deruelle, T. Dolezel and J. Katz, Phys. Rev. D\textbf{63}, 083513 (2001); hep-th/0010215v2.
\bibitem{lrr1}See \cite{lrr} for an extensive list of reference.
\bibitem{dgp}G. Dvali, G. Gabadadze and M. Porrati, Phys. Lett. B \textbf{485}, 208 (2000).
\bibitem{deffayet}C. Deffayet, Phys. Lett. B \textbf{502}, 199 (2001); C. Deffayet, G. Dvali and G. Gabadadze, Phys. Rev. D \textbf{65}, 044023 (2002).
\bibitem{mannheim}P. D. Mannheim, Phys. Rev. D \textbf{63}, 024018 (2000).
\bibitem{constcurv}See Ref. 3 above, p. 306.
\bibitem{andrianopoli}L. Andrianopoli, M. Derix, G. W. Gibbons, C. Herdeiro, A. Santambrogio and A. Van Proeyen, Class. Quant. Grav. \textbf{17}, 1875 (2000); hep-th/9912049.
\bibitem{shiromizu}T. Shiromizu, K. Maeda and M. Sasaki, Phys. Rev. D \textbf{62}, 024012 (2000).
\bibitem{israel}W. Israel, Nuovo Cimento Soc. Ital. Fis. B \textbf{44}, 1 (1966); Errata, \textbf{48}, 463 (1967).
\bibitem{ida}D. Ida, JHEP, \textbf{09} 014 (2000); S. Mukhoyama, T. Shiromizu and K. Maeda, Phys. Rev. D \textbf{62}, 024028 (2000).
\bibitem{maartens}R. Maartens, D. Wands, B. A. Bassett and I. P. C. Heard, Phys. Rev. D \textbf{62}, 041301 (2000).
\bibitem{pal}S. Pal, gr-qc /0806.2505v2.
\bibitem{durrer}R. Durrer and P. Kocian, Class. Quant. Grav. \textbf{21}, 2127 (2004).
\bibitem{tsujikawa}S. Tsujikawa and A. R. Liddle, astro-ph/0312162; C. Ringeval, T. Boehm and R. Durrer, hep-th/0307100.
\bibitem{maia1}M. D. Maia and E. M. Monte, astro-ph/0404370.
\bibitem{maia2}M. D. Maia, E. M. Monte, J. M. F. Maia and J. S. Alcaniz, astro-ph/0403072v2.
\bibitem{dahia1}F. Dahia and C. Romero, J. Math. Phys. \textbf{43}, 3097 (2002) and references therein.
\bibitem{anderson1}E. Anderson and J. E. Lidsey, Class. Quant. Grav. \textbf{18}, 4831 (2001); E. Anderson, F. Dahia, J. E. Lidsey and C. Romero, J. Math. Phys. \textbf{44}, 5108 (2003).
\bibitem{dahia2}F. Dahia and C. Romero, Class. Quant. Grav. \textbf{21}, 927 (2004).
\bibitem{felder}G. N. Felder, A. Frolov and L. Kofman, Class. Quant. Grav. \textbf{19}, 2983 (2002).
\bibitem{romero}C. Romero and F. Dahia, Int. J. Mod. Phys. A \textbf{20}, 2237 (2005).
\bibitem{chatillon}N. Chatillon, C. Macesanu and M. Trodden, Phys. Rev. D \textbf{74}, 124004 (2006); R. Koley and S. Kar, Class. Quant. Grav. \textbf{24}, 79 (2007).
\bibitem{anderson2}E. Anderson and R. Tavakol, gr-qc/0509055v2.
\bibitem{dahia3}F. Dahia, L. F. P. da Silva, C. Romero and R. Tavakol, J. Math. Phys. \textbf{48} 072501 (2007).
\bibitem{dahia4}F. Dahia, C. Romero, L. F. P. da Silva and R. Tavakol, 0711.1297 [gr-qc].
\bibitem{dahia5}F. Dahia and C. Romero, gr-qc/0702011v2.
\bibitem{maartens1}R. Maartens, Phys. Rev. D \textbf{62}, 084023(2000); hep-th/0004166; gr-qc/0101059.
\bibitem{anderson3}E. Anderson and R. Tavakol, gr-qc/0305013v3.
\bibitem{israelit}M. Israelit, Found. Phys. \textbf{35}, 1725 (2005); O. Arias, R. Cardenas and I. Quiros, Nucl. Phys. B \textbf{643}, 187 (2002); N. Barbosa-Cendejas and A. Herrera-Aguilar, Phys. Rev. D \textbf{73}, 084022 (2006).
\bibitem{dahia6}F. Dahia, G. A. T. Gomez and C. Romero, 0711.2754v1 [gr-qc].
\end{thebibliography}
\end{document}